\documentclass[12pt]{article}
\usepackage{graphics}
\input epsf

\newcommand{\abs}[1]{\left| #1 \right|}

\newcommand{\be}{\begin{eqnarray}}
\newcommand{\ben}{\begin{eqnarray}\nonumber}
\newcommand{\ee}{\end{eqnarray}}

\begin{document}
\title{
\vskip 1cm
Landscape Implications of Extended Higgs Models
}
\author{L. Clavelli\footnote{lclavell@bama.ua.edu}\\
Department of Physics and Astronomy\\
University of Alabama\\
Tuscaloosa AL 35487\\ }
\maketitle
\begin{abstract}
     From several points of view it is strongly suggested that the current
universe is unstable and will ultimately decay to one that is exactly
supersymmetric (SUSY).  The possibility that atoms and molecules form in this
future universe requires that the degenerate electron/selectron mass is
non-zero and hence that electroweak symmetry breaking (EWSB) survives the
phase
transition to exact SUSY.  However, the minimal supersymmetric standard 
model (MSSM) and several of its extensions have no EWSB in the SUSY limit.  
Among the extended Higgs models that have been discussed one stands out in
this regard.  The Higgs sector that is revealed at the Large Hadron Collider
(LHC) will therefore have implications for the future universe.  
We also address the question as to whether the transition to the
exact SUSY phase with EWSB is exothermic.
\end{abstract}


\section{\bf Introduction}
\setcounter{equation}{0}
    From the observation of a small positive vacuum energy in our world
whereas its early stages had a much higher vacuum energy,  from the
persistent indications of supersymmetry with zero vacuum energy
in the simplest manifestations
of string theory and from the dynamical connection between worlds of
differing vacuum energy in string theory,  it seems quite likely
that the present universe is intrinsically unstable.
See for example Ref.\,\cite{Giddings}.
Ultimately the universe should undergo a phase transition to a 
supersymmetric ground state.  This final phase might be a
supersymmetric anti-deSitter
world of negative vacuum energy but, apart from predicting an
ultimate big crunch, such a world would share many of the properties
of the zero vacuum energy SUSY world discussed in this article.  
We do need to assume that there is no state of massively negative
vacuum energy since the universe might rapidly fall into such
an state, if available, and would then collapse on a microscopic
time scale.  This assumption is not inconsistent with experiment
nor with any unique prediction of string theory.

Once in one of these 
SUSY states the universe should never return to a deSitter state
\cite{Linde}.  
     Other things being equal, a universe that was supersymmetric
from the beginning would probably not generate sufficient structure
to give rise to life. For this reason or for others it has been stated
\cite{Bousso} that life would be impossible in a SUSY universe.
On the other hand, if an already evolved universe became supersymmetric
through vacuum decay,
it is possible, given favorable values of the parameters of the theory, 
that life could re-evolve in the SUSY background.
The primary properties of
such a SUSY universe where all particles are degenerate with their
SUSY partners derive from a weakening of the Pauli exclusion principle
\cite{future}.  It seems that atoms and molecules could exist in such a
world \cite{CL} providing the common particle/sparticle masses in the
SUSY phase were non-zero.

In the Schroedinger
equation and in its relativistic Dirac counterpart, all atomic
and molecular
energy levels are proportional to the electron/selectron mass.
In the limit of vanishing electroweak symmetry breaking (EWSB)
atomic and molecular binding energies would, therefore, vanish
while mean radii of
atoms and molecules would tend to infinity.   
For instance, in the variational approach taken in Ref.\cite{CL},
the ground state energy and mean radius of a system with Z SUSY protons
and N SUSY electrons of mass m were found to be
\be
     E(N,Z) &=& - \frac{Nme^4}{2}\left( Z - \frac{5}{16}(N-1)\right)^2\\
     r(N,Z) &=& \frac{3}{2me^2}\left(Z - \frac{5}{16}(N-1)\right)^{-1}\quad .
\ee
One would expect that life would not arise in a world 
consisting only of elementary particles with no electromagnetic bound states.

From the point of view of the question as to whether post-transition life forms
could exist, 
the problem arises that in the Minimal Supersymmetric Standard
Model (MSSM) and in most of its extensions, electroweak symmetry
breaking vanishes in the exact SUSY limit leaving all fermions massless
as we shall discuss below.

    Similarly if the atomic masses were greater in the exact SUSY phase
than in the broken SUSY phase, energy
conservation would require an endothermic phase transition in matter from the
broken SUSY world to exact SUSY.  However, the nucleon masses are dominated by
non-perturbative QCD confinement effects and are much greater than the
light quark masses.  The masses of atoms are therefore somewhat 
insensitive to the
masses of quarks.  A small increase in the masses of SUSY 
neutrons, protons, and electrons could be compensated by the release of
vacuum energy or by the release of
excitation energy given that scalar particles are not bound by the Pauli
principle.  Of course, for isolated hydrogen or helium, this energy
source is not available and it would be complicated to consider whether
the required energy could come from nearby heavier atoms.

Therefore, for simplicity we ask if there is a theory
where the Higgs vacuum expectation values in the broken SUSY phase are
greater than or equal to those in the exact SUSY phase so that an
exothermic transition to exact SUSY could lead in a simple way to a world 
supporting atoms and molecules.  

Note that one cannot at present predict
that a future SUSY universe would or would not be such as to support atomic 
and molecular
binding.  We can, however, point to a possible determination at the Large
Hadron Collider (LHC) that the parameters of the Higgs potential are such
as to suggest an answer to this question.  The possibility of a scientific
test distinguishes the question from a purely philosophical one.

    Although we have derived some inspiration from string theory, this paper 
is exclusively an investigation of SUSY Higgs models where there are
exact SUSY as well as possible broken SUSY minima.  In these models there are no
anti-deSitter vacua so, in the absence of any experimental
constraints, we can ignore the question of whether in string
theory there are local minima with massively negative vacuum energy.

    In section II of the current paper we examine the MSSM and several
of the extended SUSY Higgs models.  We find that only one of these,
the ``nearly minimal supersymmetric standard model" (nMSSM) preserves
EWSB in the SUSY limit thus allowing non-zero masses for the fermions
and their degenerate superpartners.  

    In section III we discuss two models where spontaneous
SUSY restoration occurs preserving EWSB. These models are
analogous to other recently proposed models for meta-stable SUSY
breaking \cite{ISS} where the long lifetime of the current phase can
be accomodated.  

The meta-stability of the broken SUSY phase of the
extended Higgs models was, in fact, noted decades ago by Fayet
although one hoped then that this false vacuum could be made arbitrarily
long lived.
\cite{Fayet}. 

    Finally, conclusions are presented in section IV.

\section{\bf MSSM and Extended Higgs Models in the Exact Susy Limit}
\setcounter{equation}{0}

    The properties of several extended Higgs models have been 
the subject of much study in recent years following the pioneering
work of Fayet \cite{Fayet}.  They offer the promise
of solving the $\mu$ problem and relieving stringent experimental
constraints on the MSSM \cite{Nilles}.  The field has been
comprehensively reviewed \cite{reviews} in the past year.
For a recent summary of models with an additional Higgs singlets
see Ref.\cite{BLS}.

    The scalar potential of these models with broken SUSY consists of
F terms derivable from a superpotential, $W$,
\be
    V_F = \sum_\phi \abs {\frac{\partial W}{\partial \phi}} ^2
\ee
plus D terms and soft terms
both of which break SUSY.  The soft terms consist of scalar mass terms 
plus terms
proportional to terms in the superpotential.  The soft breaking terms
are often assumed to be consequences of a SUSY breaking mechanism in
a ``hidden sector" which interacts gravitationally or otherwise 
with the particles of our world. 

     In the exact SUSY phase, if one exists, the soft terms will 
summarily drop to zero and the D terms will disappear at the
potential minimum.

With sufficient experimentation, the LHC will be able to measure the
magnitude of soft terms and subtract them out to determine whether the
other parameters of the Lagrangian are such as to allow EWSB and
therefore molecular physics in the SUSY limit.  In this paper, therefore,
we do not consider soft terms.

In this section we investigate the exact SUSY limit of the MSSM and 
several extended models.  We drop terms in the scalar potential involving
charged fields since these have no possible vacuum expectation values
in a charge conserving theory.  
We find and investigate charge conserving
minima without prejudice as to whether or not charge violating minima
are possible.

\subsection{MSSM}

     The minimal supersymmetric standard model is defined by a 
superpotential    
\be
     W = \mu H_u \cdot H_d \quad ,
\ee
the dot product of two Higgs doublets being defined by
\be
     H_u \cdot H_d = H_u^{0}H_d^{0} - H_u^{+}H_d^{-} \quad .
\label{dotprod}
\ee

Since we do not consider the possibility that the charged Higgs fields could
acquire vacuum expectation values, we suppress any occurence of charged fields.
Thus the dot product in \ref{dotprod} is taken to be equivalent to the
product of the neutral fields. 

The F term in the scalar potential of the MSSM, restricted to neutral
fields, is
\be
     V_F = \abs{\mu} ^2 \left ( \abs{H_u}^2 + \abs{H_d}^2 \right) \quad .
\ee 
The D terms are
\be 
     V_D = \frac{g_1^2+g_2^2}{8} \left( \abs{H_d}^2 - \abs{H_u}^2 \right)^2
+ \frac{g_2^2}{2}\left( \abs{H_d}^2 \abs{H_u}^2 - \abs{H_u\cdot H_d}^2\right)^2\quad .
\label{Dterms}
\ee

Since we restrict our attention to the potential of the neutral Higgs fields, we
can discard the term in $g_2^2$.

The soft Higgs mass terms are
\be
     V_{soft} = m_d^2 \abs{H_d}^2 + m_u^2 \abs{H_u}^2 \quad .
\ee

    In the absence of the soft mass terms, the minimum of the potential
is at
\be
     <H_u> = <H_d> = 0 \quad .
\ee
     The D terms vanish at this minimum but since the Higgs fields have
zero vacuum expectation value, the electroweak symmetry breaking also
vanishes.  Thus, in the exact SUSY limit of the MSSM, there are no 
fermion masses and hence no electromagnetically bound atoms.

\subsection{NMSSM}

   The next to minimal supersymmetric standard model (NMSSM) introduces a
singlet superfield S.  The superpotential is defined by

\be
     W = \lambda S H_u \cdot H_d + \frac{\kappa}{3} S^3 \quad .
\ee

The corresponding scalar potential, restricted to neutral fields is,

\be
    V_F = \abs{\lambda S}^2 \left( \abs{H_u}^2 + \abs{H_d}^2 \right)
+ \abs{\lambda H_u H_d + \kappa S^2}^2 \quad .
\ee
The D terms are, as in the MSSM,
\be 
     V_D = \frac{g_1^2+g_2^2}{8} \left( \abs{H_d}^2 - \abs{H_u}^2 \right)^2
+ \frac{g_2^2}{2}\left( \abs{H_d}^2 \abs{H_u}^2 - \abs{H_u \cdot H_d}^2\right)^2\quad .
\ee 
The soft terms are
\be
     V_{soft} = m_d^2 \abs{H_d}^2 + m_u^2 \abs{H_u}^2 + m_S^2 \abs{S}^2 
  + \left(A_s \lambda S H_u\cdot H_d + \frac{\kappa}{3}A_\kappa S^3 + h.c.\right)
\quad .
\ee
   As the soft terms go to zero, the potential becomes symmetric in $H_u$ and
$H_d$ so that
\be
      <H_u> = <H_d> = v_0 \quad .
\ee
This ensures that the D terms vanish at the minimum.
The minimum also defines a vacuum expectation $S_0$ of the S field.
\be
      <S> = S_0 \quad .
\ee
Since the potential is positive definite or zero, any field configuration
where the potential vanishes is necessarily a true minimum.  Thus the
configuration
\be
     S_0 = v_0 = 0
\ee
is a true supersymmetric minimum.  Since $V_F$ is a sum of positive definite
terms this is the only SUSY minimum and it fails to provide for EWSB.

\subsection{UMSSM}

    This minimal SUSY model with an extra U(1) is defined by the
superpotential
\be
      W = \lambda S H_u\cdot H_d\quad .
\ee
together with an extra U(1) gauge symmetry coupling to the two Higgs
and to the S field with charges $Q_u$, $Q_d$, and $Q_S$.
The F terms in the scalar potential are
\be
    V_F = \lambda^2 \left( \abs{H_u H_d}^2 + \abs{S}^2 (\abs{H_u}^2 + \abs{H_d}^2)\right)\quad .
\ee
The D terms are as in the MSSM plus terms from the additional U(1):
\be
    V_D = V_{D,MSSM} + \frac{g_U^2}{2}\left( Q_{H_d}\abs{H_d}^2
+Q_{H_u}\abs{H_u}^2+ Q_S\abs{S}^2\right)^2 \quad.
\ee
In exact SUSY, the D terms must vanish at the minimum of the potential.

The symmetry of the superpotential implies that $Q_S+Q_{H_u}+Q_{H_d}=0$.
The vanishing of $V_F$ and $V_D$ at the SUSY minimum requires in
any case that the positive definite term in $|H_u H_d|^2$
vanishes at the minimum and, therefore, $v_0=0$, i.e. no EWSB in the SUSY 
limit.

\subsection{nMSSM}

    The nearly minimal supersymmetric standard model (nMSSM) is defined by
the superpotential
\be
    W = \lambda S \left(  H_u\cdot H_d - v^2 \right)\quad .
\ee
where, again, $S$ is an electroweak singlet superfield.

    The F terms in the scalar potential are then
\be
    V_F = \lambda^2 \left( S^2 ( \abs{H_u}^2 + \abs{H_d}^2 ) + (\abs{H_u H_d} - v^2)^2 \right) \quad .
\ee
The D terms are the same as in the MSSM in eq.\ref{Dterms}, vanishing at the 
generic potential minima:
\be
     <H_u>=<H_d>=v_0 \quad , \quad <S> = S_0 \quad .
\ee
The soft terms correspond to mass terms for the scalars plus terms proportional
to terms in the superpotential.  These soft terms, possibly governed by
SUSY breaking in a hidden sector, vanish in the SUSY limit.
In this exact SUSY limit, setting the soft terms to zero, the extrema
are defined by 
\be
    \left.\frac{\partial V}{\partial S}\right|_{_0}= 2 S_0 v_0^2 = 0
\ee
and
\be
        \left.\frac{\partial V}{\partial H_u}\right|_{_0} = v_0(v_0^2 - v^2+S_0^2) =
        0 \quad .
\ee 
In the SUSY limit the absolute minimum of the scalar potential is at
\be
        v_0 = v \quad , \quad S_0 = 0\quad .
\ee
Thus for the nMSSM in the SUSY limit there is vanishing vacuum energy and
with a broken electroweak symmetry ($v_0 \neq 0$).

All of the models of this section have been extensively studied in
\cite{Fayet} and subsequent papers.  It was, in fact, noted there that
the nMSSM would allow an exact SUSY with EWSB.  This observation
acquires relevance in the current context of a final transition to
the exact SUSY phase.
In the following section we will study two toy models with a meta-stable
broken SUSY
minimum decaying to an exact SUSY with EWSB.  These are both
generalizations of the nMSSM (and the other extended Higgs models).  

\section{\bf Meta-stable Models with EWSB in the Exact Susy Limit}
\setcounter{equation}{0}

    In this section we consider first a model coupled
through singlet Higgs fields to a mirror world (hidden sector)
indicated by tildes and then secondly a model with no mirror symmetry.
 
\subsection{meta-stable model with mirror symmetry}

    The Higgs superpotential is taken to be
\be
       W = \lambda \left( S (H_u\cdot H_d - v^2) + \tilde{S}(\tilde{H}_u\cdot \tilde{H}_d - v^2) + \mu_0  S \tilde{S} \right) \quad .
\ee
The mirror Higgs fields do not couple directly to normal matter and
they may or may not be coupled to mirror matter.  As we shall show, the
corresponding scalar potential has minima with broken and unbroken SUSY.
The neutral field F terms in the scalar potential take the form
\ben
    V_F &=& \lambda^2 \left( \abs{H_u H_d - v^2+ \mu_0 \tilde{S}}^2 +  \abs{S}^2(\abs{H_u}^2 +\abs{H_d}^2)\right.\\  
&+& \left.\abs{\tilde{H_u}\tilde{H_d} - v^2 + \mu_0 S}^2 + \abs{\tilde{S}}^2(\abs{\tilde{H}_u}^2 +\abs{\tilde{H}_d}^2)\right) \quad .
\ee
The D terms in the potential are as in the mirrored nMSSM.
\ben
     V_D &=& \frac{g_1^2+g_2^2}{8} \left( \abs{H_d}^2 - \abs{H_u}^2 \right)^2
+ \frac{g_2^2}{2}\left( \abs{H_d}^2 \abs{H_u}^2 - \abs{H_u H_d}^2\right)
\\
&+&\frac{g_1^2+g_2^2}{8} \left( \abs{{\tilde H_d}}^2 - 
\abs{{\tilde H_u}}^2 \right)^2
+ \frac{g_2^2}{2}\left( \abs{\tilde{H_d}}^2 \abs{\tilde{H_u}}^2 - 
\abs{\tilde{H_u}\tilde{H_d}}^2\right)\quad .
\ee

   The vacuum expectation values of the Higgs, to be determined by the
minimization conditions, are
\be
    <H_u>&=& <H_d> = v_0\\
    <\tilde{H}_u> &=& <\tilde{H}_d> =\tilde{v}_0 \\
    <S> &=& S_0\\
    <\tilde{S}>&=& \tilde{S}_0 \quad .
\ee
The minimization conditions are
\be
     \left.\frac{1}{\lambda^2} \frac{\partial V_F}{\partial \tilde{S}}\right|_{_0}=
0 = \mu_0 (v_0^2-v^2 + \mu_0 \tilde{S}_0) + 2 \tilde{v}_0^2 \tilde{S}_0
\label{cond1}
\ee
\be
     \left.\frac{1}{\lambda^2} \frac{\partial V_F}{\partial H_u}\right|_{_0} =
0 = v_0  (v_0^2-v^2 + \mu_0 \tilde{S}_0 + S_0^2 )
\label{cond2}
\ee
\be
     \left.\frac{1}{\lambda^2} \frac{\partial V_F}{\partial S}\right|_{_0} =
0 = \mu_0 (\tilde{v}_0^2-v^2 + \mu_0 S_0) + 2 v_0^2 S_0
\label{cond3}
\ee
\be
    \left. \frac{1}{\lambda^2} \frac{\partial V_F}{\partial \tilde{H}_u}\right|_{_0} =
0 = \tilde{v}_0  (\tilde{v}_0^2-v^2 + \mu_0 S_0 + \tilde{S}_0^2 )
\label{cond4}
\ee

The most obvious solution is\\
{\bf Solution 1: $v_0 = \tilde{v}_0 = v \quad , \quad S_0 = \tilde{S}_0 = 0$}\\
This is the ground state of the model and corresponds to an exact supersymmetry
( vanishing vacuum energy), and with EWSB ($v_0 \ne 0$). 

A broken SUSY solution with, however, no EWSB, lies at
{\bf $v_0 = \tilde{v}_0 = 0 \quad , \quad S_0 = \tilde{S}_0 = 
\frac{v^2}{\mu_0}\quad .$}\\

{\noindent\bf A SUSY breaking minimum} can be found 
with non-zero parameters $v_0, \tilde{v}_0, S_0, \tilde{S}_0$.
We can see from eqs.\,\ref{cond1},\ref{cond2},\ref{cond3}, and \ref{cond4} that
\be
    \tilde{S}_0 &=& \frac{\mu_0}{2 \tilde{v}_0^2}S_0^2\\
    S_0 &=& \frac{\mu_0}{2 v_0^2}\tilde{S}_0^2 \quad .
\ee
The solution with nonvanishing vevs for the two singlet Higgs are
\be
      S_0 &=& \frac{2 v_0 \tilde{v}_0}{\mu_0}{(\frac{\tilde{v}_0}{v_0})}^{1/3}\\
      \tilde{S}_0 &=& \frac{2 v_0 \tilde{v}_0}{\mu_0}{(\frac{v_0}{\tilde{v}_0})}^{1/3}\quad .
\ee
                                     
Substituting these into the minimization conditions gives complementary
cubic equations for $v_0^2$ and $\tilde{v}_0^2$: 
\be
   (v_0^2 - v^2)^3 &=& - 8(1 + \frac{2 \tilde{v}_0^2}{\mu_0^2})^3 \tilde{v}_0^2 v_0^4\\
   (\tilde{v}_0^2 - v^2)^3 &=& - 8(1 + \frac{2 v_0^2}{\mu_0^2})^3 v_0^2 \tilde{v}_0^4 \quad .
\label{cubiceqs}
\ee

These have the sole solution:
\be
    v_0^2 = \tilde{v}_0^2 = \frac{3 \mu_0^2}{8}\left(\sqrt{1 + \frac{16 v^2}{9\mu_0^2}}-1 \right) \quad .
\ee
The vev's have an upper limit 
\be
      v_0 = \tilde{v}_0 = v /\sqrt{3} \quad .
\ee
Thus, the degenerate supermultiplets are 
heavier in the exact SUSY phase than the fermions in the broken SUSY phase.  This would
make the transition to exact SUSY endothermic.

\subsection{ A meta-stable Model without Mirror Symmetry}

   A simpler model without the mirror symmetry of the previous
section is the following.

Consider the most general renormalizable Higgs superpotential with a pair of
doublets and a single extra Higgs singlet supermultiplet \cite{Fayet}:

\be
    W = \lambda \left( S (H_u\cdot H_d - v^2) + \frac{\lambda^\prime}{3} S^3
           + \frac{\mu_0}{2}S^2 \right) \quad .
\label{general}
\ee

If $v$ and $\mu_0$ are taken to vanish this is the superpotential of the $NMSSM$.
If $\lambda^\prime$ and $\mu_0$ are absent, this is the $nMSSM$. If all of
$\lambda^\prime$,$\quad \mu_0$, and $v$ vanish and an additional U(1) gauge
interaction is introduced, this becomes the UMSSM.  
  
Since, the superpotential of eq.\,\ref{general}
contains all the previously mentioned models as special cases, we refer to it
simply as the Singlet Extended Susy Higgs Model (SESHM).

The F terms in the scalar potential are
\be
   V_F = \lambda^2 \left( \abs{H_u\cdot H_d - v^2 + \lambda^\prime S^2 + \mu_0 S}^2   + \abs{S}^2 (\abs{H_u}^2 + \abs{H_d}^2)\right) \quad .
\label{VF}
\ee  
The D terms are as in the MSSM, eq.\,\ref{Dterms}, vanishing at the minima of the
potential.  The soft terms are
\ben
    V_{soft} &=& m_{H_u}^2 \abs{H_u}^2 + m_{H_d}^2 \abs{H_d}^2 
+ m_S^2 \abs{S}^2 \\
     &+& (A_s \lambda S H_u \cdot H_d  
     + A_1 \lambda v S + A_2 \lambda \mu_0 S^2 + A_3 \lambda \lambda^\prime S^3 + h.c.)\quad .
\label{Vsoft}
\ee
It seems that little if any work has been done on models with a non-zero $\mu_0$.
This parameter can be set to zero, though with some loss of generality, by imposing a
discrete symmetry under $S \rightarrow -S$ in the scalar potential.
Note that, although continuous symmetries can also be seen in the superpotential,
the scalar potential can exhibit additional discrete symmetries.

In this paper we will neglect the soft SUSY breaking terms.
They will, in any case, vanish if SUSY breaking disappears but they would, if present, 
quantitatively affect the analysis
of the broken SUSY phase and, therefore, of the inter-phase relationships that we are interested in.  

The crucial point for the present paper is that a sufficiently detailed experimental study of the Higgs potential at the LHC can isolate the soft terms and determine whether the remaining terms are such as to allow EWSB in the SUSY limit.

Similarly, in the present analysis, we will ignore phases in the Higgs sector.  
In later work we will go beyond these
toy models to incorporate phases and soft terms.
Phases could be interesting from the point of 
view of CP violation and could also affect the inter-phase relationships.

Ignoring the soft Higgs mass terms, the symmetry of the scalar potential
guarantees that the vevs of the two Higgs are equal and at this symmetry
point the D terms vanish.  

The conditions for an extremum of the scalar potential F terms are
\be
     \left.\frac{1}{\lambda^2} \frac{\partial V_F}{\partial S}\right|_{_0}=
0 = (2 \lambda^\prime S_0 + \mu_0) (v_0^2-v^2 + \mu_0 S_0 + \lambda^\prime S_0^2) + 2 v_0^2 S_0
\label{conda}
\ee
\be
     \left.\frac{1}{\lambda^2} \frac{\partial V_F}{\partial H_u}\right|_{_0} =
0 = v_0  (v_0^2-v^2 + \mu_0 S_0 + (\lambda^\prime +1) S_0^2 )\quad .
\label{condb}
\ee

Since the potential is positive definite or zero, any localized solution
of eqs.\,\ref{conda} and \ref{condb} with vanishing vacuum energy, $V_F(0)$, 
is necessarily a minimum.

The most obvious solution is\\
{\bf Solution 1: $v_0 = v \quad , \quad S_0 = 0$}\\
This is one of the two grounds state of the model and corresponds to an exact supersymmetry
( vanishing vacuum energy) with EWSB ($v_0 \ne 0$). 

A second solution is\\
{\bf Solution 2: $v_0 = 0 \quad , \quad S_0 = \frac{-\mu_0 \pm \sqrt{\mu_0^2+
4 \lambda^\prime v^2}}{2 \lambda^\prime}$}\\
This solution is also supersymmetric with a vanishing vacuum energy at the
minimum but with no EWSB.

A third solution with SUSY breaking but no EWSB is\\
{\bf Solution 3: $v_0 = 0 \quad , \quad S_0 = \frac{-\mu_0}{2 \lambda^\prime}$}\\

If both $v_0$ and $S_0$ are non-zero, we find a {\bf solution 4} with SUSY breaking
plus EWSB.  The conditions become
\be
    2 \lambda^\prime S_0^2 + \mu_0 S_0 - 2 v_0^2 = 0
\label{S01a}
\ee
\be
    (\lambda^\prime + 1) S_0^2 + \mu_0 S_0 + v_0^2 - v^2 = 0\quad .
\label{S01b}
\ee
One can combine these conditions to find
\be
    (2 \lambda^\prime +1) S_0^2 + \frac{3}{2}\mu_0 S_0 - v^2 = 0
\label{S02a}
\ee
and
\be 
     (\lambda^\prime -1) S_0^2 - (3 v_0^2 - v^2) = 0\quad .
\label{S02b}
\ee
This latter equation then predicts $v_0^2 > v^2/3$ if $\lambda^\prime > 1 $
and $v_0^2 < v^2/3$ if $\lambda^\prime < 1 $.

\begin{figure}[htbp]
\begin{center}
\epsfxsize= 3in 
\leavevmode
\epsfbox{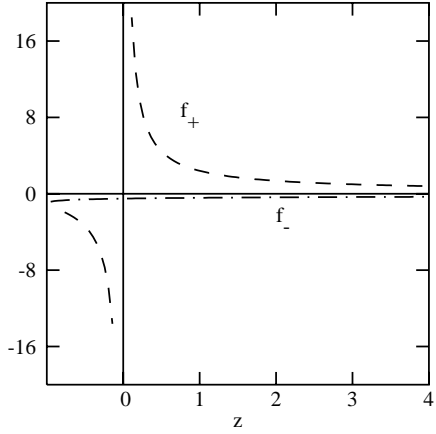}
\end{center}
\caption{The function $f_{+}(z)$ (dashed line) and the function $f_{-}(z)$ (dot-dashed line)
shown as functions of $z$.  Both functions go to $-1$ at $z=-1$.}
\label{fz}
\end{figure}

In solving equation \ref{S02a} it is convenient to define the variable
\be
      z = \frac{16 v^2 (2 \lambda^\prime + 1)}{9 \mu_0^2}
\ee
in terms of which
\be
       \mu_0 S_0{_\pm} = - \frac{4 v^2}{3 }f_{\pm}(z)
\label{mu0S0}
\ee  
with
\be
      f_{\pm} = (1 \pm \sqrt{1+z})/z \quad.
\ee
The functions $f_\pm(z)$ are shown in figure\,\ref{fz}.

From eqs. \ref{S01a} and \ref{S02a} we have
\be
     v_0^2 = \frac{1}{2 \lambda^\prime + 1}(\lambda^\prime v^2 - \mu_0 S_0 (\lambda^\prime -1)/2)\quad .
\label{v0}
\ee  
Together these give $S_0$ and $v_0$ as a function of the parameters, $\lambda^\prime$ and $\mu_0$.

In a future paper we would like to perform a complete analysis of the
structure of minima and maxima of the scalar potential. Here we would only like to show that at least one 
true SUSY breaking minimum does exist.

The conditions for such an extremum to be a true minimum as opposed to a maximum or a saddle point
is that all the eigenvalues of the mass squared matrix be positive.  Neglecting phases, 
the mass squared matrix in the space of $H_u$, $H_d$, and $S$ as
obtained from the second derivatives of the scalar potential is
\ben
   M^2 = \left| \begin{array}{ccc}    \quad  \alpha+\zeta \quad & \quad \gamma-\zeta \quad & \quad \delta \quad\\
        \quad  \gamma-\zeta  \quad & \quad \alpha + \zeta \quad & \quad \delta \quad\\
        \quad \delta  \quad      &  \quad \delta  \quad       & \quad \beta \quad \end{array} \right|
\ee
where
\be
     \alpha = \left.\frac{\partial^2 V_F}{\partial H_u^2}\right|_{0} = 2 \lambda^2 (v_0^2 + S_0^2)
\ee
\be
     \gamma = \left.\frac{\partial^2 V_F}{\partial H_u \partial H_d}\right|_{0} = 2 \lambda^2 (v_0^2 - S_0^2)
\ee
\be 
     \delta = \left.\frac{\partial^2 V_F}{\partial H_u \partial S}\right|_{0} = 2 \lambda^2 v_0 \left(2
 (\lambda^\prime + 1) S_0 + \mu_0 \right)
\ee
\be
     \beta = \left.\frac{\partial^2 V_F}{\partial S^2}\right|_{0} = 2 \lambda^2 \left( 6 {\lambda^\prime}^2 {S_0}^2 + 6 \lambda^\prime S_0 \mu_0+ 2 \lambda^\prime (v_0^2 - v^2) + \mu_0^2 + 2 v_0^2 \right)
\ee
and
\be
     \zeta =  \left.\frac{\partial^2 V_D}{\partial H_u^2}\right|_{0} = (g_1^2+g_2^2) v_0^2  \quad .
\ee    
The eigenvalues of the mass squared matrix satisfy
\be
     m_1^2 = 4 \lambda^2 S_0^2 + 2 \zeta \quad ,
\ee
\be
     m_2^2 + m_3^2 = \alpha + \beta + \gamma \quad ,
\ee
and 
\be
     m_2^2 m_3^2 = \beta (\alpha + \gamma) - 2 \delta^2 \quad .
\ee
The first squared mass is positive definite and corresponds to the eigenvector
\be
     \Psi_1 = \frac{H_u-H_d}{\sqrt 2} \quad .
\ee

The positivity of $m_2^2$ and $m_3^2$ puts constraints on the parameter space of
$\lambda^\prime$ and $\mu_0$,  namely
\be
    m_2^2+m_3^2 = 2 \lambda^2 \left( 2 \lambda^\prime v^2 + \mu_0^2 + 
\mu_0 S_0 (\lambda^\prime +2)\right) > 0
\ee
and
\be
    m_2^2 m_3^2 = - (2 \lambda^2)^2 4 v_0^2 v^2 (1 + f_\pm(z)) > 0 \quad .
\label{prod}
\ee
From eq.\,\ref{prod} and figure\,\ref{fz}, one can see that positive Higgs masses requires  
negative $z$ and the choice of the positive root $f_{+}$.  
This in turn requires that $\lambda^\prime$ be
sufficiently negative.  Numerically we find that we must have $\lambda^\prime < -2$.

\begin{figure}[htbp]
\begin{center}
\epsfxsize= 3in 
\leavevmode
\epsfbox{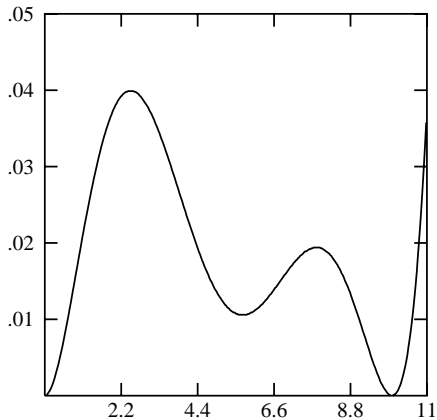}
\end{center}
\caption{The effective potential as a function of the magnitude of the
doublet Higgs field showing qualitatively the false vacuum of broken SUSY with EWSB 
and the two degenerate vacua of exact SUSY.  The numerical scales are given in arbitrary 
units.  The extremum with broken SUSY but without EWSB is not shown.}
\label{potential}
\end{figure}

With the assumptions of the current paper (neglect of
soft Higgs masses and neglect of phases) the Higgs potential is
described by figure\,\ref{potential}.  Thus, if there is no change in yukawa couplings,
the transition from the broken SUSY phase to the exact SUSY phase with EWSB is
endothermic. In this paper we do not discuss solution three which, depending on the parameters may also
be a true, local, SUSY-breaking minimum but one with no EWSB.  
Particle masses are products of Yukawa couplings and Higgs vevs.  To relate the particle masses in
separate phases we must know whether and by how much the yukawa couplings change in the transition.
Such changes of couplings do not occur in the usual treatment of lagrangian Higgs models
so we neglect them 
but they may occur in string theory (with presently undetermined magnitudes).

The value of the Higgs potential at the broken SUSY minimum is
\be
    V_F(0) = \lambda^2 S_0^2 (S_0^2 + 2 v_0^2) \quad .
\ee
$V_F(0)$ gives a contribution to the dark energy in the broken SUSY phase.
Other contributions come from the mass splitting of fermions and sfermions
and, perhaps, from compactification and thermal effects.  The algebraic sum of 
these various contributions could be equated with the measured vacuum
energy.

\section{\bf Conclusions}
    The primary result of the current paper 
is that, neglecting phases, a true broken SUSY minimum exists over a 
finite region of the parameter space without assuming the presence
of additional soft mass terms.  This requires a non-zero $\mu_0$ 
parameter which was not part of previous analyses. Furthermore,
while the MSSM, the NMSSM, and the UMSSM do not support a phase transition 
to an exact SUSY with
non-zero fermion masses (EWSB),  the SESHM with non-zero $v$ does have 
EWSB as required for
atomic and molecular structure in the SUSY phase.
Thus, if the LHC finds a non-zero $v$ in the scalar potential,
the possibility is open that SUSY atoms and molecules will
form after the expected phase transition to exact SUSY although,
at present, it seems that an additional energy input would be
required to permit the transition to proceed.  

     We present two toy models which have a meta-stable
broken SUSY phase and an exact SUSY phase with EWSB.  In these models
one or more singlet Higgs with vacuum quantum numbers play the role of
an inflaton field mediating the transition between phases of differing
vacuum energy.  

     A second SUSY ground state (solution 2) has no EWSB.  In this
ground state, matter would be permanently ionized with no possibility of
atomic structure. 

      It is not within the scope of the current paper to examine the
phenomenology of the broken SUSY phase (solution 4) beyond the
basic requirements discussed above so much remains to be done.  
Previous work without allowing for the contribution of the $\mu_0$ parameter
has been
discussed in the reviews \cite{reviews}\cite{BLS} of the extended Higgs models
but it remains to be seen whether the currently allowed parameter space
or that to be revealed in future LHC experiments is consistent with $v>0$.
By measuring the $\lambda$, $\lambda^\prime$, and $\mu_0$ parameters and
the Higgs masses, the LHC can also determine whether soft Higgs masses are
needed in the broken SUSY phase.  

At the present level of analysis, neglecting phases, soft SUSY breaking terms, 
and the possibility of
the yukawa couplings decreasing in the transition, the common particle/sparticle
mass in the exact SUSY phase is greater than the particle mass in the broken SUSY phase.
Thus the transition would require extra energy input to proceed.  This
energy could come from the energy (primarily nuclear) stored in the Pauli towers.  
Otherwise, the transition would necessarily be to the SUSY phase without EWSB
(solution 2).  An interesting question for future study would be whether
further extending the Higgs potential with additional Higgs multiplets
or considering non-zero phases in the vev's
could allow a broken SUSY phase with $v_0 > v$.

{\bf Acknowledgements}

    This work was supported in part by the US Department of Energy under
grant DE-FG02-96ER-40967.   We are grateful for the
hospitality of the Institute for Fundamental Theory at the University
of Florida during the winter quarter of 2007 when most of the results
of this paper were extracted.  Some observations of Salah Nasri at
the University of Florida were key to the directions taken in this work.  
We also acknowledge the hospitality of the INFN at the
University of Bologna and several discussions on the topics of this paper
with the particle theory group there in the spring of 2007.

\end{document}